\documentclass[a4paper,12pt,oneside]{article}
\usepackage[english]{babel}
\usepackage{amsmath}
\usepackage{amssymb}
\usepackage{epsf}
\usepackage{graphicx}
\usepackage[usenames]{color}

\usepackage{tocenter}
\FromMargins[f]{15mm}{15mm}{20mm}{20mm}

\graphicspath{{img/}}

\title{The effect  of  the electron-beam parameter spread
on microwave generation in a three-cavity axial vircator}
\author{E.A. Gurnevich, P.V. Molchanov \\
Research Institute for Nuclear Problems, Belarusian State
University,\\
11 Bobruiskaya Str., Minsk 220030, Belarus;\\
e-mail:molchanov$@$inp.bsu.by; genichgurn$@$gmail.com}
\date{}

\begin{document}
\maketitle

\begin{abstract}
 The behavior  of the generation
efficiency and radiation spectrum of a three-cavity axial vircator
versus the radius of the injected electron beam, its impedance and
energy homogeneity is studied. This paper establishes that for
each geometry of a three-cavity resonator, there exist  the
optimal (maximizing the generation efficiency) values of these
parameters and determines the range within  which they may vary
without losing the efficiency of generation.
\end{abstract}
\maketitle

\section{Introduction}

Power sources of electromagnetic radiation  with virtual cathodes
(vircators) have become a subject of intense study \cite{Vircator,
HPM} lately. The capability to produce gigawatt power level
microwave pulses makes the vircator-based sources promising
devices for such applications as plasma physics and techniques,
high energy density physics, accelerator physics and techniques,
and radar and communication technologies. High-current electron
accelerators demonstrate unique capabilities  as seed sources for
powering such microwave oscillators. The power of high-current
electron beams achievable at present is  $10\div 100$~GW, while
the corresponding radiation power of the vircator is as high as
tens of gigawatts \cite{HPM,Mesyats}.

The main advantages of vircator-based oscillators are high output
 power of microwaves, the possibility to generate without external
 guiding fields, and compactness.
  Their chief disadvantages lie in low efficiency, instability of the generation frequency
during the pulse length and its poor reproducibility  in
experimental series.
 The typical generation efficiency achievable in experiments is
 only $1\div2$\%.

 Resonant electrodynamic systems placed into the
 region of VC formation offer a means to increase the
 generation efficiency and stabilize the frequency of microwave radiation  \cite{Virc_cont}.
This approach is  embodied in a three-cavity axial vircator
(Fig.~\ref{fig:geometry}). Under the electron beam parameters
(energy of 630-700 keV and current of about 21 kA) reported in
\cite{China}, the three-cavity axial vircator realizes a steady
single-frequency ($f=4.1$~GHz) generation regime with output power
as high as $\sim 1$~GW  and generation efficiency of 6.6\%.

The authors of  \cite{INP} performed a PIC simulation of a similar
three-cavity vircator with the generation frequency from 3 to 4
GHz in several geometries. It was demonstrated that the efficiency
of such oscillators can be rather high ($\sim 5$~\% and higher)
even at quite small beam energies (under 500 keV). But the
simulation described in \cite{INP} was quite idealized: the
parameters of the electron beam injected into the resonator were
considered to be set rigorously, and the beam was assumed to be
homogeneous and monochromatic.
However, in actual experiments with three-cavity resonators, the
electron beam is generated in a high-current diode with an
explosive-emission cathode. In this case, the absolute values of
the beam energy and current vary significantly during the voltage
pulse, as well as their ratio, i.e.,  the beam impedance (due to
the expansion of the cathode plasma); different  dynamical effects
cause  the beam radius to vary (it does not always equal to the
radius of the emitting surface of the cathode).
Moreover a certain energy spread always exists in the  initial
beam (it is not monochromatic). Thus it is practical  to know what
effect the imperfection of the beam and/or the slight variation of
its parameters may have on the  efficiency and spectrum of
generation. This paper studies how  the generation efficiency and
the radiation spectrum depend on  the electron-beam radius,
impedance, and  energy spread in each of the three designs of a
three-cavity vircator suggested in \cite{INP}.

\section{Simulation results}

\begin{figure}[htbp]
\begin{center}
\includegraphics[scale=0.6]{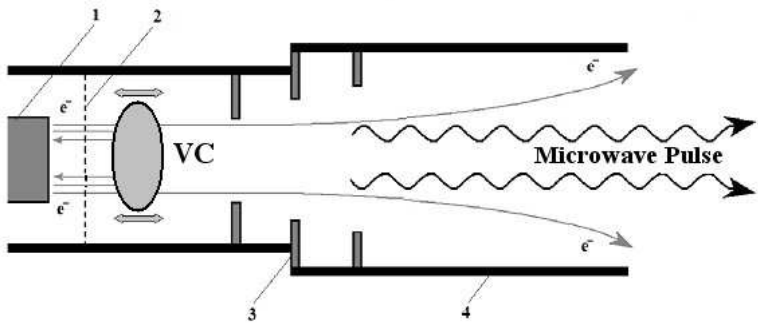}
\end{center}
\caption{Scheme of a three-cavity vircator.} \label{fig:geometry}
\end{figure}

To simulate the electron beam dynamics in an axial vircator, we
used the particle-in-cell method, realized in a free 2.5D PIC code
XOOPIC \cite{XOO}. The oscillator is shown schematically in
Fig.~\ref{fig:geometry}, where  1 is the emitting cathode, 2 is
the anode mesh, 3 are the resonator sections of the vircator, and
4 is the cylindrical output drift tube.
The electron beam of specified configuration is injected into the
first resonator cavity directly, and so all the processess
occurring in the cathode-anode gap are neglected. For all vircator
designs considered here, the electron beam energy was assumed to
be constant and equal to E=450 keV \cite{INP} (generation in such
vircators at various beam energies was studied in \cite{EG,PM}).

In the selected scheme of a three-cavity vircator, the virtual
cathode oscillates in the first resonant cavity (where the
beam-microwave energy conversion occurs), and  we shall
distinguish between the resonators by its  dimensions $L_1$
 and $r_1$ (length and radius). The parameters of the considered resonators \cite{INP} are listed
 in Table~\ref{tab:resonators}.

\begin{table}[htp]
  \centering
  \begin{tabular}{|p{5cm}|p{2cm}|p{2cm}|p{4cm}|}
    \hline
    Resonator   &   $L_1$, mm   &   $r_1$, mm   & Radiation frequency  $f$, GHz \\
    \hline
    \#1     &   60  &   50  &   3.2\\
    \hline
    \#2     &   54  &   45  &   3.6\\
    \hline
    \#3     &   48  &   45  &   3.7\\
    \hline
 \end{tabular}
\caption{Resonator designs}
 \label{tab:resonators}
\end{table}

\begin{figure}[htbp]
\begin{center}
\includegraphics[scale=1.0]{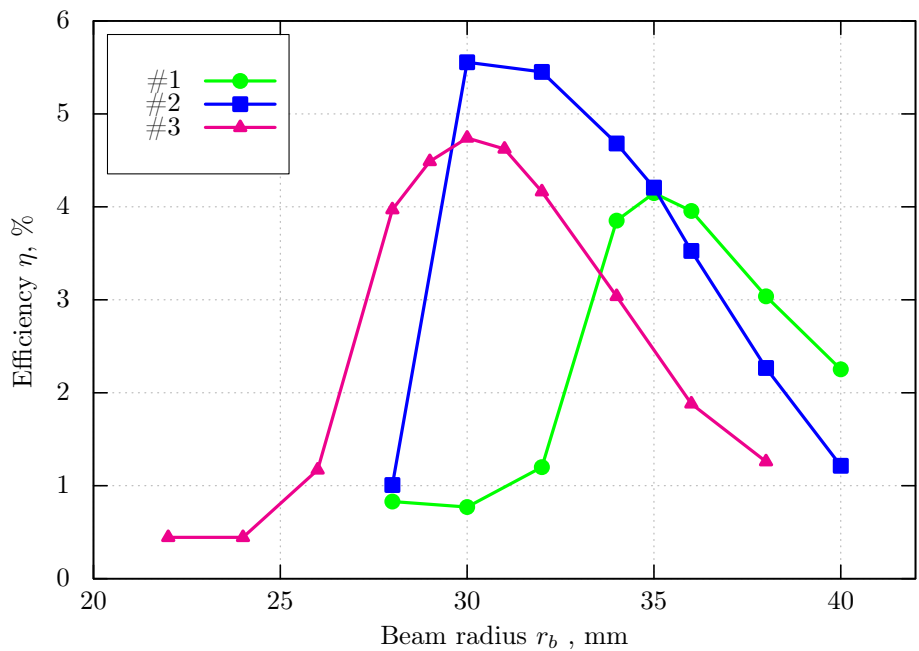}
\end{center}
\caption{Generation efficiency as a function of the electron beam
radius for three vircator designs. The beam energy and current are
450 keV and 15 kA, respectively.} \label{fig:eff_rb2}
\end{figure}

\begin{figure}[htbp]
\begin{center}
\includegraphics[scale=1.0]{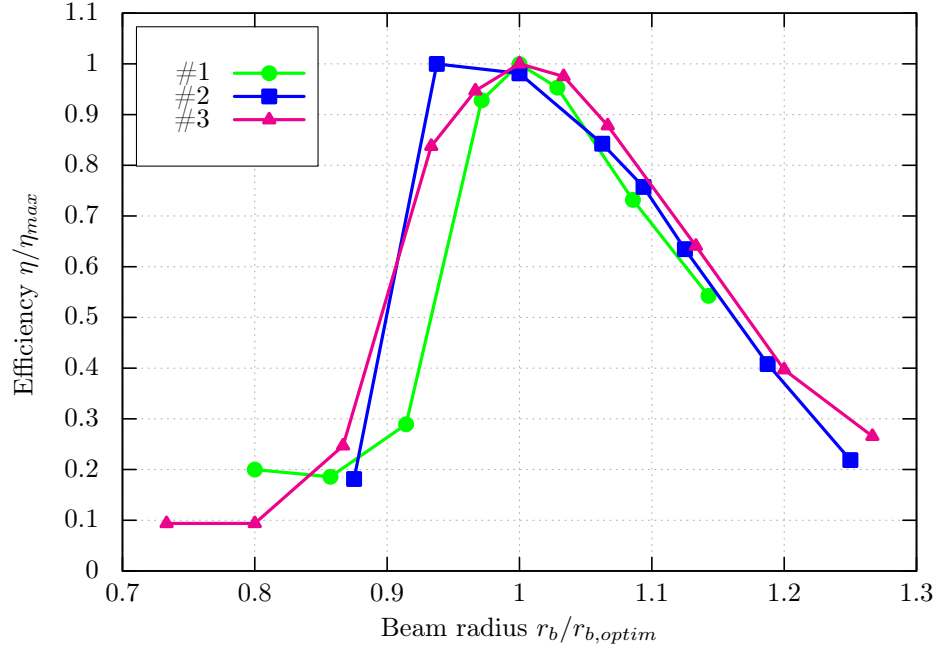}
\end{center}
\caption{Generation efficiency as a function of electron-beam
radius: $r_b$ and $\eta$ are normalized to their optimal values;
the beam energy and current are 450 keV and 15 kA, respectively.}
\label{fig:eff_rb4}
\end{figure}

\begin{figure}[htbp]
\begin{center}
\includegraphics[scale=1.0]{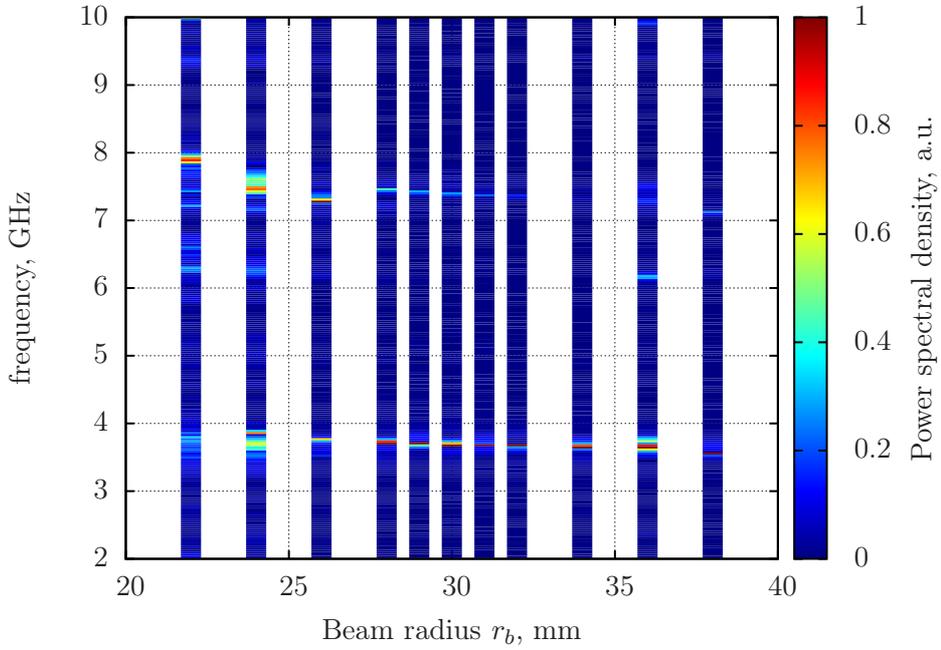}
\end{center}
\caption{Radiation spectra for vircator  \#3 at different values
of the beam radius.} \label{fig:spec_rb}
\end{figure}

We first study the generation characteristics as a function of the
electron-beam radius at constant beam current of 15 kA, the same
for all three designs. The simulation results are given in Figs.~\ref{fig:eff_rb2}-\ref{fig:spec_rb}.

As is seen, there is an optimal radius of the electron beam for
each design of the three-cavity vircator at which the maximum
generation efficiency is achieved.
Let us note that all three dependencies are similar, which is
obvious from Fig.~\ref{fig:eff_rb4}, and for this reason we can
suppose that $\eta$ will reveal a similar dependence on $r_b$ in
other possible designs of three-cavity resonators, too.
It should also be mentioned that as the beam radius is increased
(compared to its optimal value), the generation efficiency reduces
slower than in the case when the beam radius diminishes. Thus a
10\% increase in the beam radius  leads to a 30\%  decrease in the
generation efficiency, while a 10\%  decrease in the beam radius
leads to a 50-80\% drop in the radiation efficiency, approaching
the radiation efficiency of a conventional axial vircator (without
resonant cavities).
This can be explained, in particular, by the fact that with
decreasing radius of the beam (at constant current and energy) the
major radiation frequency in the vircator increases sharply
because the beam  plasma frequency grows (the generation starts to
occur at the second harmonic), which is accompanied by spectral
broadening (Fig.~\ref{fig:spec_rb}).  The drop in the efficiency
with growing beam radius is related to the  decrease in the
beam-current supercriticality, and finally to the vanish of the
conditions for VC formation.

\begin{figure}[htbp]
\begin{center}
\includegraphics[scale=1.0]{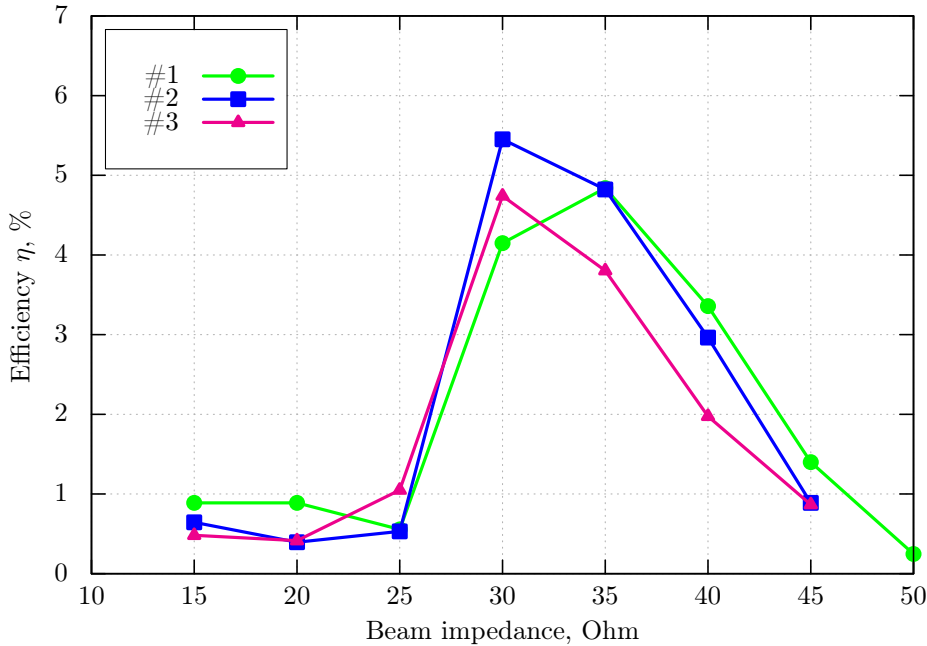}
\end{center}
\caption{Generation efficiency as a function of  electron beam
impedance for three vircator designs; the energy of the beam
electrons equals 450 keV. } \label{fig:eff_Z}
\end{figure}

The optimal values of the electron beam radius: $r_{b1}=35$~mm,
$r_{b2}=32$~mm, and $r_{b3}=30$~mm, obtained from the simulation
results were used for further computations.
The efficiency dependence on the beam impedance in three-cavity
vircators was studied by simulating the system behavior at
different currents of the injected beam (the energy was set to be
450 keV and did not vary). The maximum efficiency $\eta \sim 5$~\%
corresponds to the impedance $Z\sim 30 \div 35$~Ohm for all
studied vircator designs and drops to 1\% whenever the impedance
changes by 10~Ohm (Fig.~\ref{fig:eff_Z}).
This circumstance is crucial for practical realization of the
three-cavity vircator, because in actual high-current diodes with
explosive-emission cathodes, the impedance is constantly
decreasing through cathode plasma expansion.

\begin{figure}[htbp]
\begin{center}
\includegraphics[scale=1.0]{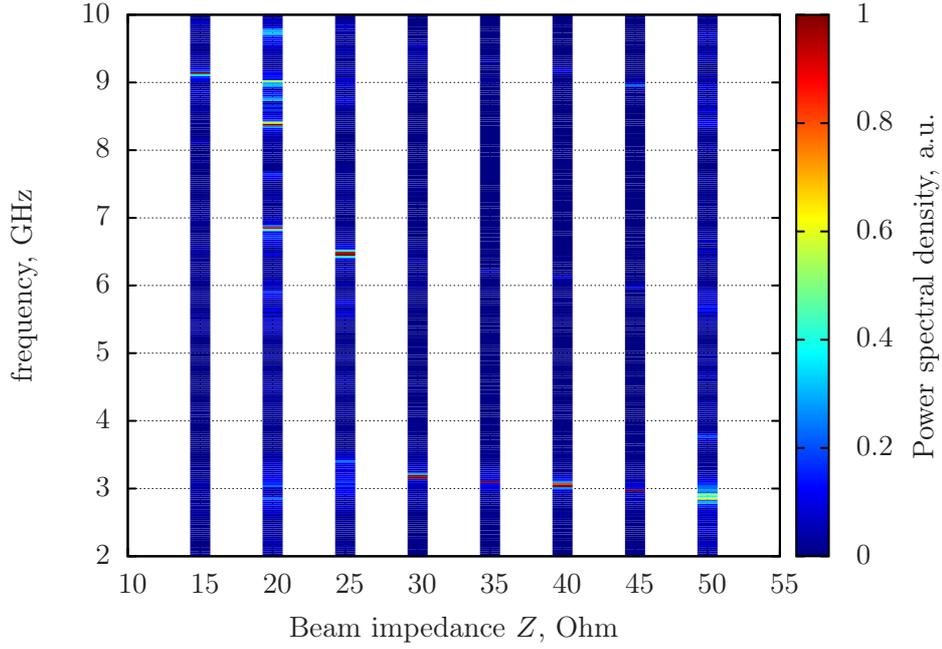}
\end{center}
\caption{Radiation spectra for vircator  \#1 at different values
of the beam impedance (the energy of the beam electrons equals 450
keV ).} \label{fig:spec_Z}
\end{figure}

Plasma frequency increases with diminishing beam impedance, as
well as with decreasing beam radius,   and the moment comes when
the main generation frequency shifts to the range of higher
harmonics (Fig.~\ref{fig:spec_Z}).
  In fact, with decreasing beam impedance
(i.e., with increasing beam current) the radiation efficiency
grows monotonously  from the moment when the VC formation becomes
possible until the generation is single-mode and occurs at the
main  working frequency of the resonator. The generation
efficiency drops sharply as the generation shifts to the range of
high frequencies.

\begin{figure}[htbp]
\begin{center}
\includegraphics[scale=1.0]{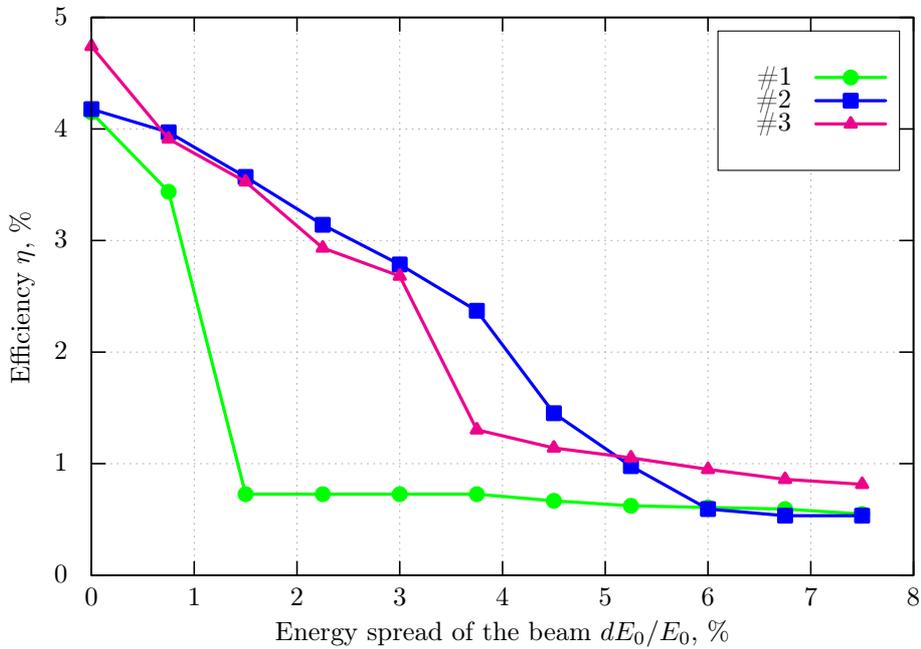}
\end{center}
\caption{Generation efficiency as a function of  thermal spread in
energies of  beam electrons for three vircator designs. The beam
energy and current are 450 keV and 15 kA, respectively.}
 \label{fig:eff_dE}
\end{figure}

\begin{figure}[htbp]
\begin{center}
\includegraphics[scale=1.0]{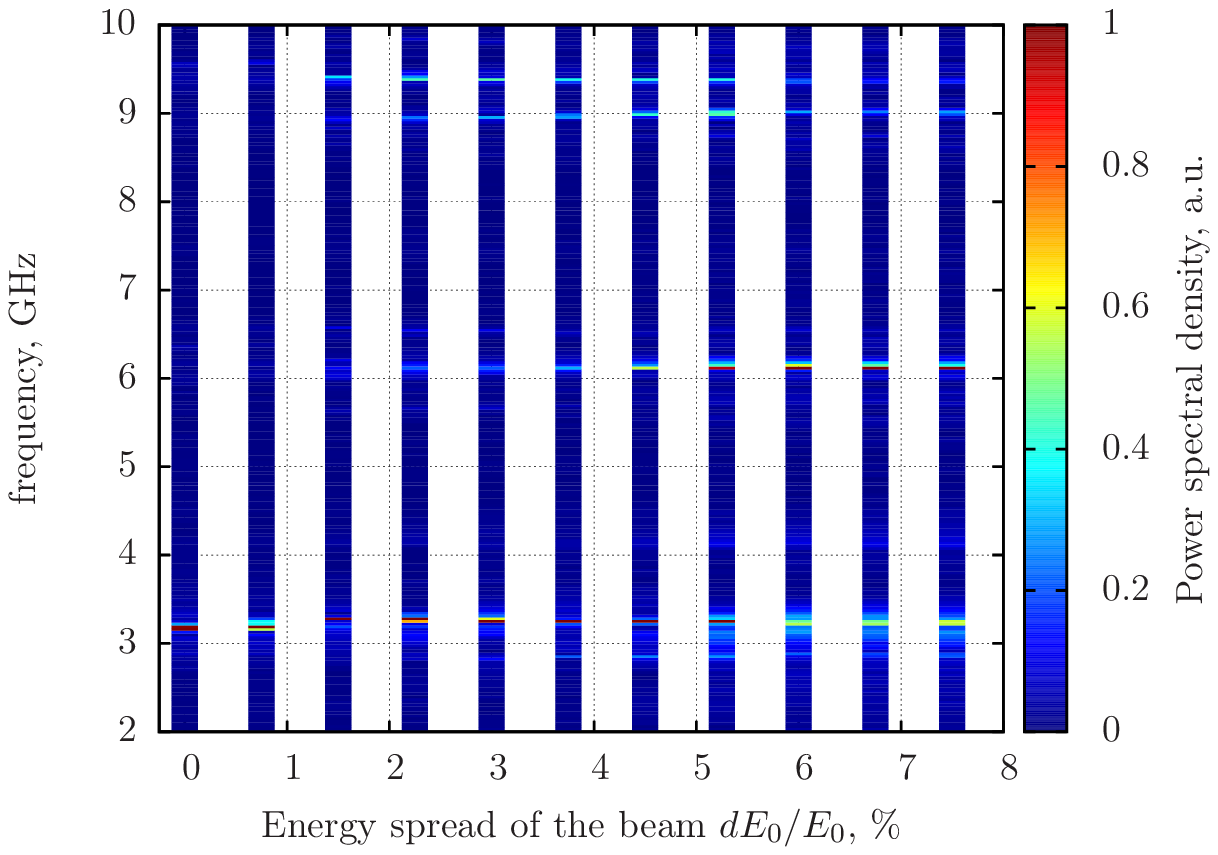}
\end{center}
\caption{Radiation spectra for vircator  \#1 as a function of beam
energy spread.} \label{fig:spec_dE_evg1}
\end{figure}

\begin{figure}[htbp]
\begin{center}
\includegraphics[scale=1.0]{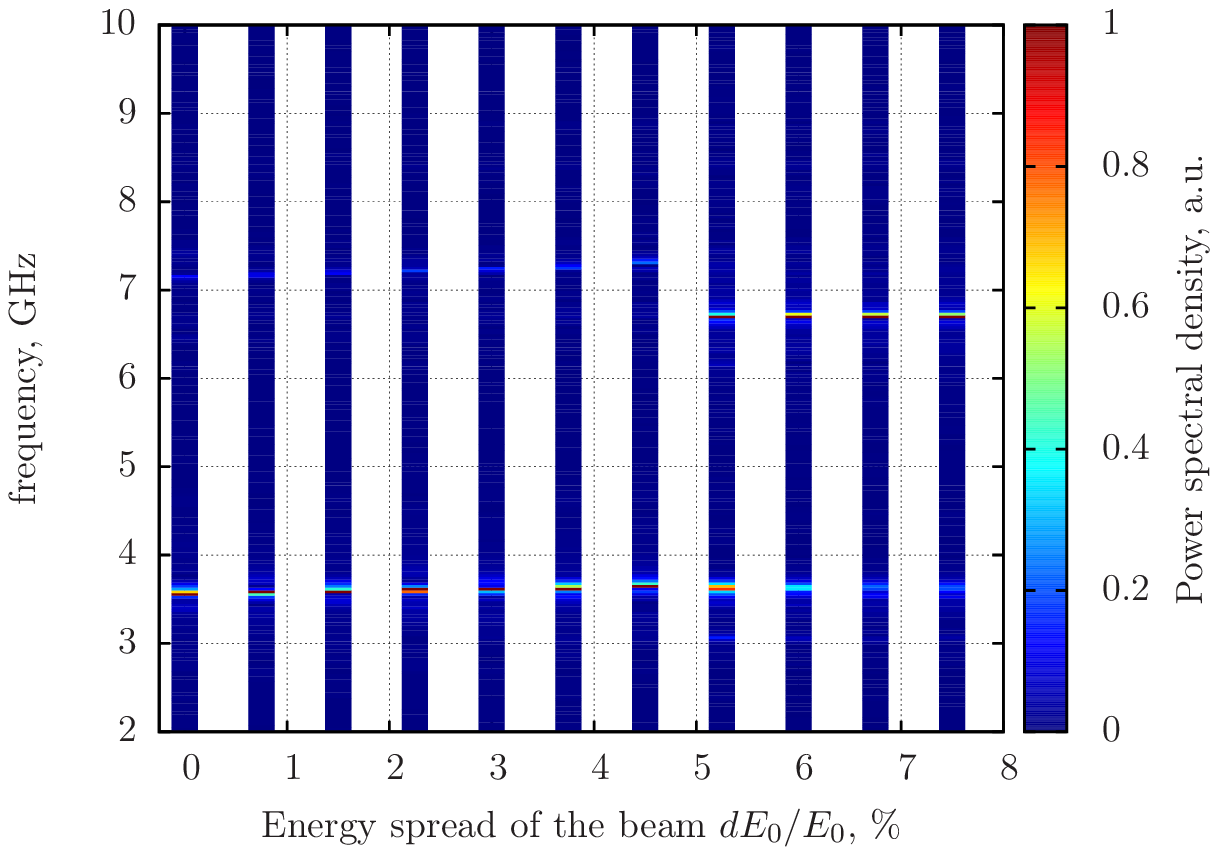}
\end{center}
\caption{Radiation spectra for vircator  \#2 as a function of beam
energy spread.} \label{fig:spec_dE_pav2}
\end{figure}

Thermal spread in the energies of  beam electrons can also
noticeably diminish the generation efficiency $\eta$ (see also
\cite{DUB,Jiang1995}). By way of example,  Fig.~\ref{fig:eff_dE}
shows $\eta$ vs $dE_0/E_0$, where $E_0=450$~keV and $dE_0
(=3/2kT)$ is the average  kinetic energy of thermal motion of beam
electrons. When the spread is as small as 5\% (or even 1.5\% for
vircator   \#1 ), the efficiency drops to $\sim 1$~\%, which is
typical of axial vircators without resonant cavities.
Figures~\ref{fig:spec_dE_evg1} and \ref{fig:spec_dE_pav2} show
that in this case, the radiation spectra broaden, and the
amplitudes of higher harmonics in them are increased.

\section{Conclusion}
This paper presents a numerical simulation of microwave generation
in a three-cavity axial vircator  with the electron beam energy of
450 keV and the frequency of $3\div 4$~GHz. This paper also
analyzes the dependence of the generation efficiency and spectrum
on the radius of the injected beam, its impedance and the presence
of the energy spread. The relations obtained here enable us to
conclude that microwave generation in the considered vircator has
a resonant character. It has been  found that for each geometry of
the three-cavity resonator, there is an optimal radius of the
electron beam  that provides  maximum generation efficiency (as
high as 5\% and even more) that can drop dramatically (to 1\%)
when the beam radius differs by 10\% from the optimal value. It is
determined that the optimal values of the electron beam impedance
for a three-cavity vircator  lie in the range from 30 to 35~Ohm.
It has also been established that as the electron beam energy
spread is enhanced, the generation efficiency falls off fairly
rapidly and drops to a fraction of the previous value even when
the spread is as small as 5\%.

\end{document}